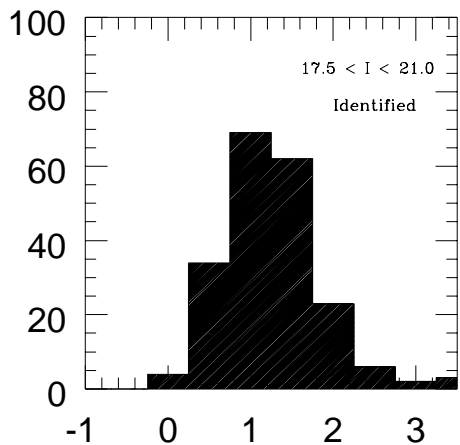 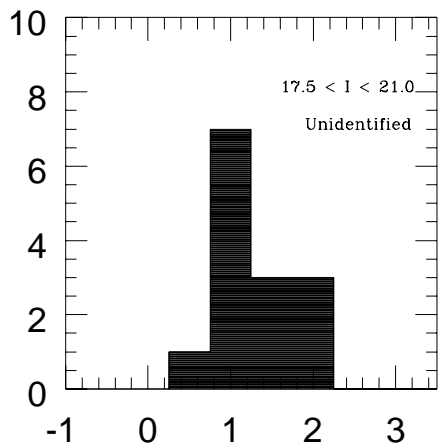 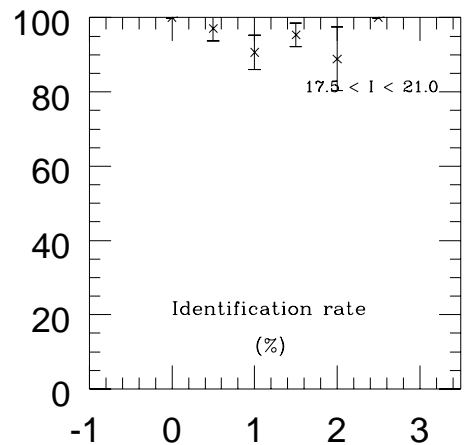
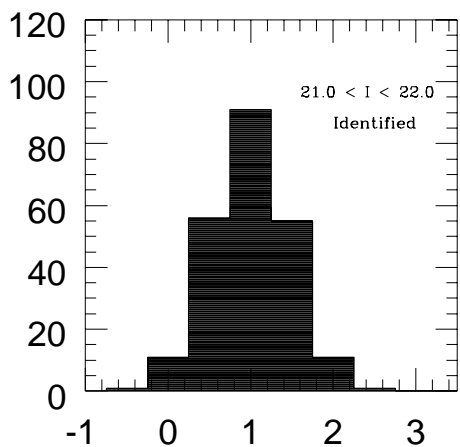 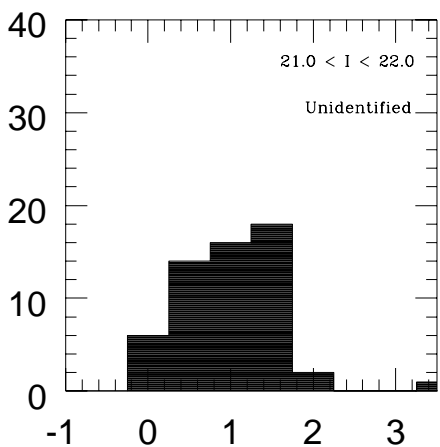 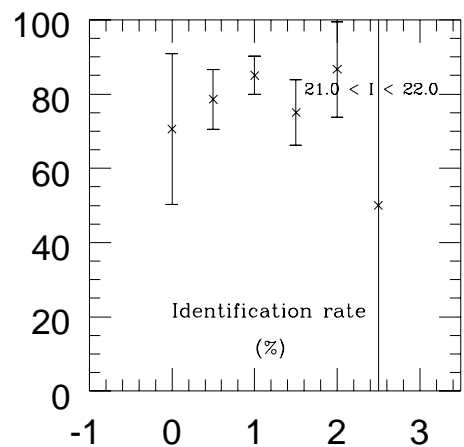
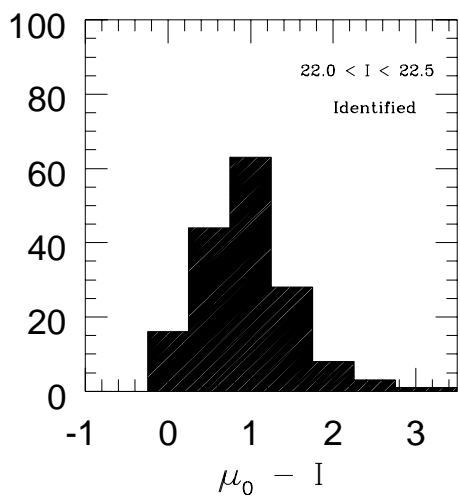 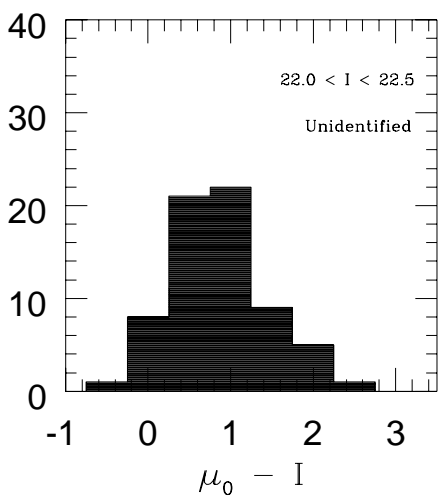 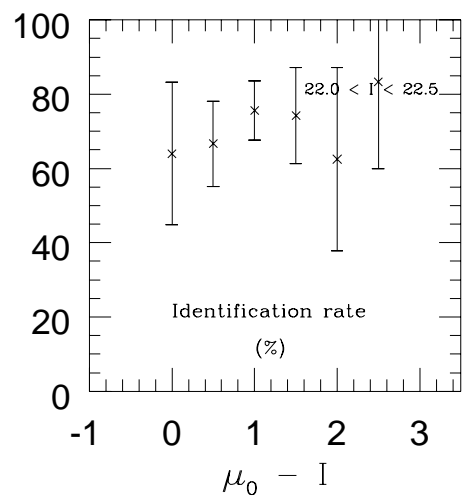

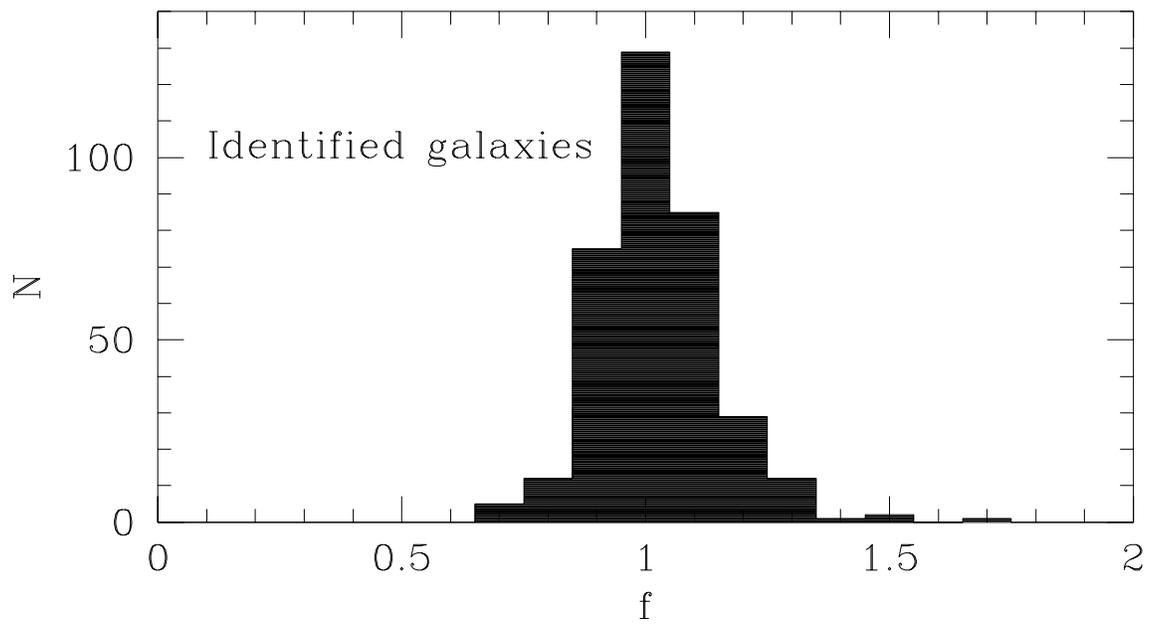
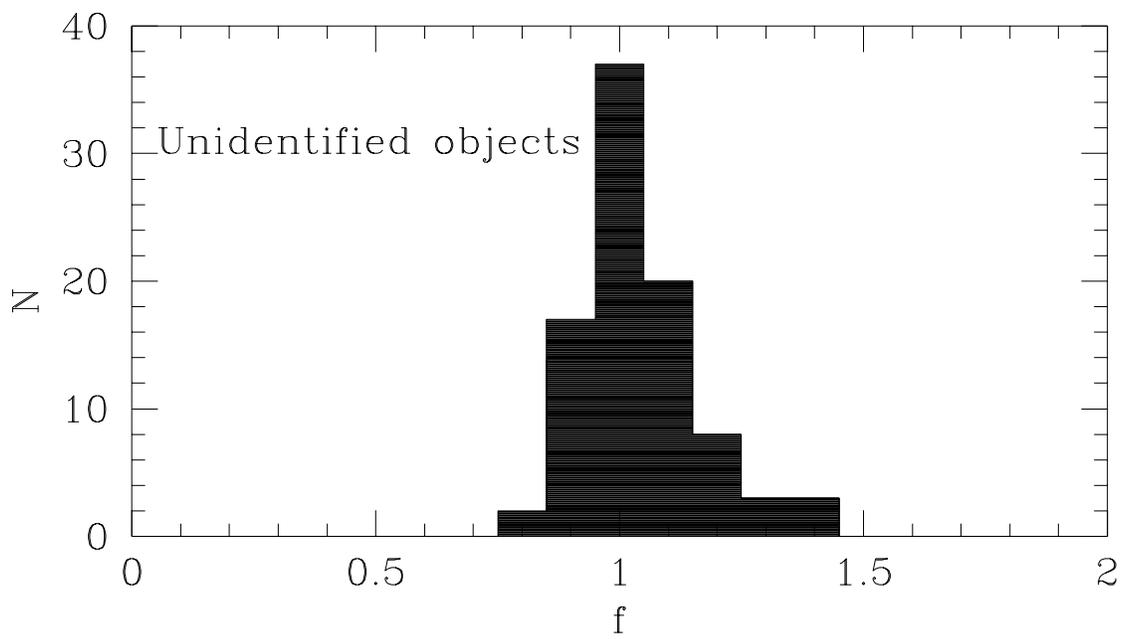

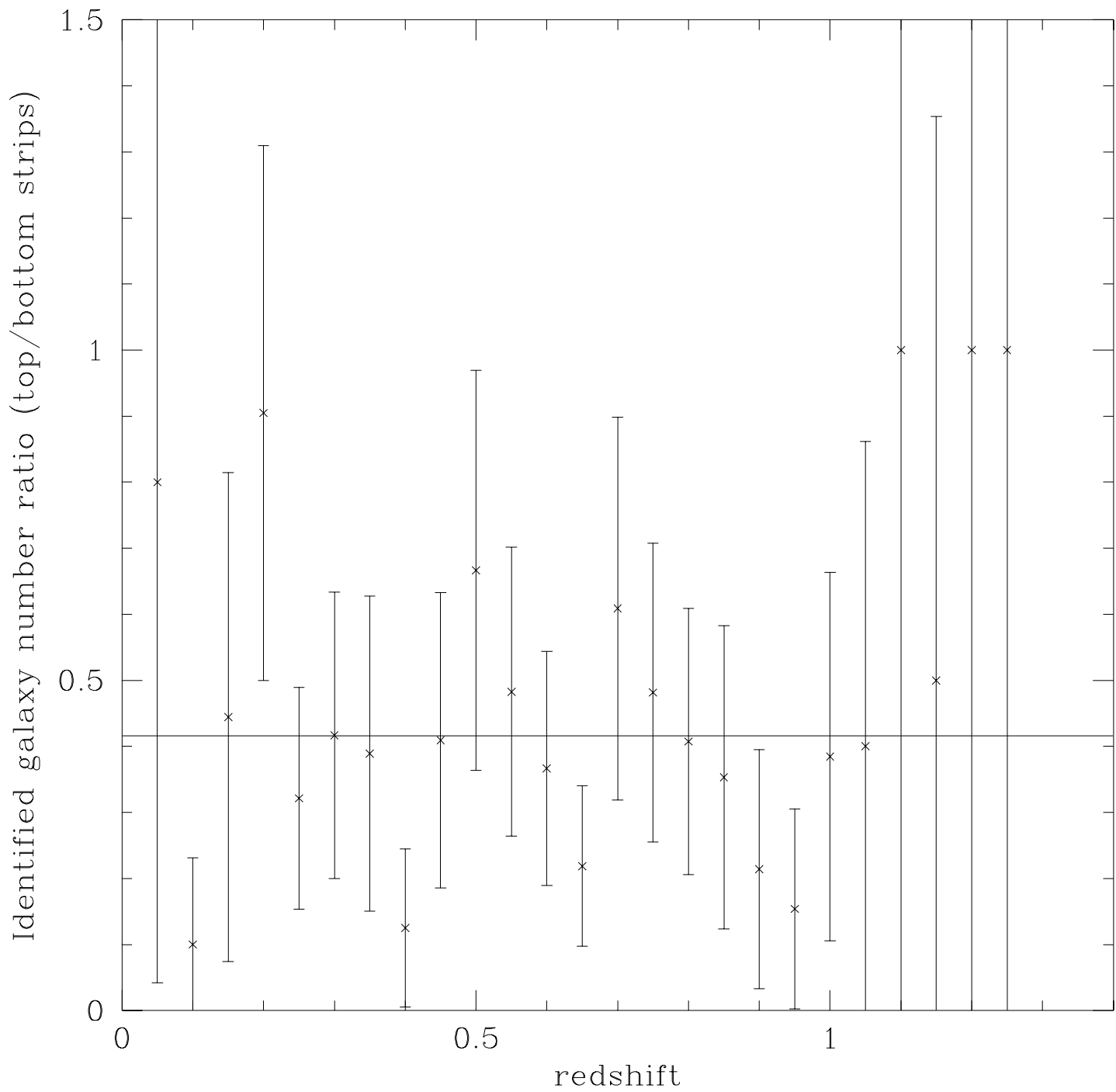

# The Canada-France Redshift Survey IV: Spectroscopic selection effects and 0300+00 field spectroscopic data

Francois Hammer[1]
DAEC, Observatoire de Paris-Meudon, 92195 Meudon, France

David Crampton[1]
Dominion Astrophysical Observatory, National Research Council of Canada, Victoria, Canada

Olivier Le Fèvre[1]
DAEC, Observatoire de Paris-Meudon, 92195 Meudon, France

and

Simon J. Lilly[1]
Department of Astronomy, University of Toronto, Toronto, Canada M5S 1A7

## ABSTRACT

Possible surface brightness selection effects in the redshift catalogs of the Canada-France Redshift Survey are investigated through comparisons of subsamples of the data. Our analyses demonstrate that the securing of redshifts is independent of possible biases arising from surface brightness effects and/or differing galaxy morphologies and orientations. The unusual geometry of the mask designs for our spectroscopic observations also do not produce any significant bias. There is however a bias at the highest and lowest redshifts, especially for absorption-line galaxies at z>1 and z<0.2, due to the adopted spectral range (4250Å to 8500Å). Apart from the latter, we conclude that our sample of identified galaxies is an unbiased subsample of the original photometric catalogue and is essentially limited by I-band flux density ($17.5 \leq I_{AB} \leq 22.5$). Finally, spectroscopic data for 273 objects in the 0300+00 CFRS field are presented.

*Subject headings:* galaxies: distances and redshifts

## 1. Introduction

---

The CFRS project has been a major effort to gather a complete sample of ∼1000 spectra of very faint field objects ($I_{AB}$ ≤22.5) (The AB magnitude system is used throughout this paper; $I_{AB} = I + 0.48$). The main goals of this survey and a discussion of the establishment of reliable photometric catalogs made prior to the spectroscopy are given by Lilly et al. (1995a; CFRS I). Our deep imaging provides a sample of galaxies with isophotal $I_{AB}$ ≤22.5, which is substantially complete for central surface brightness as faint as $\mu(I)$=24.5. Le Fèvre et al. (1995; CFRS II) describe the fundamental methodology adopted for our spectroscopy, and Lilly et al. (1995b;CFRS III) discuss the accuracy and reliability of our redshift measurements. Here we present the last set of spectroscopic data of the CFRS (in the 0300+00 field) and discuss possible selection effects related to our spectroscopy, which might affect the statistical properties of the resulting sample.

Several aspects of the methodology of our spectroscopic program, could conceivably have lead to differences in the ease with which spectroscopic identifications were secured. Two obvious possibilities are related to object morphology and surface brightness, since peculiar galaxy morphologies could cause a smaller fraction of their light to pass through the spectrograph slit, and the unidentified objects might be expected to predominantly be low surface brightness galaxies or very elongated galactic disks orientated perpendicular to the slits. In addition, the unusual mask geometry (3 strips of slits per mask, see CFRS II) could potentially lead to a redshift bias since the bright zero order contamination from objects in the adjacent strips tends to affect a particular range of wavelengths in the spectra. In this paper we investigate surface brigthness selection effect in Section 2, selection effect related to the galaxy orientation relatively to the slit in Section 3 and selection effect coming from our mask geometry in Section 4, and conclude that they are all in fact insignificant.

In Section 5, we discuss one selection effect that is, however, almost certainly present in the sample. This arises from the finite wavelength range of the spectra which makes the spectroscopic identification of absorption line objects at very high and very low redshifts difficult.

Finally, we present in Section 6 the spectroscopic catalog for 273 objects in the 0300+00 survey field.

## 2. Surface brightness selection effects

Uncontrolled selection biasses against low surface brightness galaxies might cause severe biases in the determination of the luminosity function. For example, it has been suggested that the Loveday et al. (1992) local luminosity function might be biased against low surface brightness galaxies (Ferguson and McGaugh, 1995), which might affect the corresponding slope at faint luminosities. For the CFRS project, objects were selected to have isophotal $17.5 \leq I_{AB} \leq 22.5$ from very deep images. This procedure should produce an unbiased sample which includes nearly all objects with central surface brightness as faint as $\mu(I)$=24.5 (see CFRS I). Selection of the objects for spectroscopy was done without regard to the morphological properties of an object.

"Blind" techniques were used for selecting objects from the photometric catalog (i.e. without regarding their brightness and compactness), and strict criteria, independent of the object surface brightness, were applied to reject spectra for instrumental reasons (see CFRS II). We examine here whether objects for which we have failed to assign a redshift are preferentially the lowest surface brightness galaxies, for which one might expect relatively low signal-to-noise per pixel on their 2D spectra.

We have tested this hypothesis by comparing the surface brightness properties of two different samples, the first one containing the galaxies with a secure redshift identification (confidence class $\geq 2$), the second one consisting of all objects for which we failed to obtain a secure spectroscopic identification. Figure 1 shows the histograms of the distribution of the compactness of the objects in these two samples parameterized by ($\mu_{AB}(I)$-$I_{AB}$) (see CFRS I) in three different magnitude ranges. While the distributions of the two samples, identified galaxies and unidentified objects, differ slightly from one another (e.g., in the 21< $I_{AB}$ <22 range), the overall success rate in identifying a galaxy redshift is independent of the compactness. Even at the faintest magnitudes, there is no evidence for any systematic biases for or against galaxies with central surface brightness as faint as $\mu(I)=24.5$ in our final spectroscopic sample relative to the photometric one. As noted in CFRS I, virtually all the $I_{AB} \leq 22.5$ "normal" galaxies should have central surface brightnesses higher than this value.

## 3. Elongated galaxies oriented perpendicular to the slit

Another bias in the spectroscopic identifications might arise from highly elongated sources (e.g., edge-on disks) which lie perpendicular to the slit. We generate a simple test, again by comparing the properties of two samples, one containing the galaxies with a secure redshift identification (classification $\geq 2$), the other including all the objects for which we failed to secure a spectroscopic identification. The distribution of a simple geometric parameter which accounts for the object elongation as well as for the major axis orientation relative to the slit was then investigated.

Assuming $e$ is the source eccentricity and $\theta$ is the major axis orientation angle relative to the slit, the ratio of the one dimensional source section crossed by the slit to the source radius, $f = $ sqrt((1- $e^2$ cos$\theta^2$)/sqrt(1-$e^2$)), is an appropriate estimator of the orientation effect. For circular sources $f = 1$; for sources having their major axis along the slit length ($\theta$=90°) $f > 1$; and as the eccentricity and the orientation angle decreases $f < 1$. The orientation effect could potentially affect all those sources with isophotal diameters (2×$r_{28}$) larger than the slit width (1$''$75), and so $e$ and $\theta$ were measured in a homogeneous way for all such objects in the $00^h$, $03^h$ and $10^h$ fields. Figure 2 presents the distribution of the identified galaxies and unidentified objects for various values of the $f$ parameter. There is no evidence that these distributions differ significantly, implying the absence of any systematic bias for or against elongated sources in our spectroscopic sample.

## 4. Zero order contamination

The peculiar geometry of our mask design (three strips of slits, see CFRS II) led us to investigate whether this could have been introduced biases in our spectroscopic sample. CFRS II (Figure 1) shows that most of the zero orders are superimposed on the blue part of the spectra (mostly below 5500Å), and thus severely contaminate a wavelength range of 200-300Å. The zero orders contamination can cause difficulties in recognizing some of the most common features (e.g. [OII] 3727, the 4000Å break) found in galaxy spectra. This could, in turn, affect the identification of absorption-line galaxies with $z < 0.38$ or emission-line galaxies with $z < 0.48$.

Since the putative zero orders can only affect the spectra in the two lower strips of spectra, comparison of the unidentified fraction of objects with that for objects in the top row should reveal any bias. Figure 3 demonstrates that no such bias is present. Furthermore, the redshift histograms for the identified galaxies located in the top rows in the different masks is identical to that for the galaxies in the two lower strips. For instance, the fraction of galaxies with $z < 0.48$ in the top strip (65 of 174) is very similar to that for the two bottom strips (155 out of 418 galaxies). We thus conclude that there is no significant bias arising from overlapping zero orders.

## 5. Biases related to observed spectral range

The most important selection effect expected in any spectroscopic work of this kind, is caused by the unavoidable limitation of the spectral range. A relatively large spectral range was used for CFRS, from 4250Å to 8500Å (see CFRS II). Our spectra show that there are basically no features that are useful for redshift identification below 3727Å at rest (or below 4000Å at rest for an absorption-line galaxy). This is clearly a limitation of our survey, since identification of $z > 1.28$ emission-line galaxies or $z > 1$ absorption-line galaxies is virtually impossible from our spectra. A quantitative estimation of this selection effect will be presented in CFRS V.

The detection of spectral features in the blue region of our spectra is limited by the decreasing efficiency of the CCD below 4500Å rather than by our spectral wavelength limit (4250Å). It results in a reduced efficiency, especially in identifying red absorption-line galaxies at low redshift ($z < 0.2$). Indeed, in our complete sample, we have identified no such galaxies at $z < 0.15$, while 4 would be expected from the number identified at higher redshift. Although the number of objects possibly affected is small due to the small volumes accessible in our survey, the shape and errors of the derived luminosity function at low redshift could obviously be affected.

## 6. Spectroscopic data in the 0300+00 field

This field is located at $\alpha(2000)=03^h\ 02^m\ 39\overset{s}{.}5$ and $\delta(2000)=00°10'21''\!.3$. A finding chart is shown in Fig. 4 and a grey-scale reproduction showing the observed objects in Fig. 5. In this field there are 611 objects with $17.5 \leq I_{AB} \leq 22.5$, and we obtained spectra for 271 (44%) of them. We also obtained spectra of two objects at a fainter magnitude level. The statistically complete sample (see CFRS II) contains 252 objects, including 48 stars and 2 QSOs. The success rate in redshift identification (confidence class $\geq 2$) is 82.5%, the average redshift is 0.54 and the corresponding redshift distribution is shown in Figure 6.

Several observational problems (poor blue coating on the CCD, cirrus, see CFRS II) were experienced during the first observations (5 masks in October 1992, December 1992 and February 1993) of this field. These spectra were thus poorer than average and the overall redshift identification rate (with confidence class $\geq 2$) fell to $\sim 60$%. Many of these objects were reobserved with 3 new masks (6, 7 and 8), designed to include most of the unidentified objects in masks 1, 2, 4 and 5. Objects in mask 3 were abandoned since it's quality was so poor. These reobservations plus a careful analysis of the data (including co-addition of spectra of the same object observed several times) allowed us to reach an identification rate comparable to that of the other fields.

Table 1 presents the data for the 0300+000 field. The first part lists the objects which constitute our complete sample (see Paper II) and the second part contains objects classified in the supplementary catalog. The first column gives the object CFRS name, columns (2) and (3) give the 2000 coordinates, column (4) gives the isophotal I(AB) magnitude, column (5) gives the $(V-I)_{AB}$ color index, column (6) gives the compactness parameter (see Paper I), columns (7) and (8) list the object redshift (0.000 for a star and 9999 for an absence of spectroscopic identification) and the corresponding confidence classes (class 0, no identification; class = 1, poor confidence in the redshift; classes 2, 3 and 4, confidence levels higher than 85%, 99% and 100% respectively; and classes 8 and 9, single emission line objects; see CFRS II). A list of the spectral features used to determine the redshift are given in the final columns of the table.

## 7. Conclusion

This paper completes a series of three papers which present the spectroscopic data of the CFRS, and discuss the methodology and the limitations of our spectroscopic work. In this paper, possible biases are explored which might have originated from our observational methods. It is demonstrated that our strategy, adopted to optimise our multiplexing gain (i.e. three strips of slits per observing masks), has apparently not affected our identification of galaxies at moderate z, even though bright zero orders contaminate $\sim 200$Å in the blue part of many spectra. The surface brightnesses, orientations and eccentricities of the unidentified objects compared to those of the identified galaxies shows that any biases related to these properties are also insignificant. Apart from small biases at the lowest and highest redshift regimes arising from the limited wavelength range (4250Å – 8500Å) of our spectra, our final spectroscopic catalog is essentially unbiased relative to the original photometric sample from which it was selected, and is limited by one single selection criterion, the I-band flux density.

We thank the directors of the CFHT and the two allocation time committees (CFGT and CTAC) for their continuing support and encouragement. SJL's research is supported by the NSERC of Canada. We acknowledge some travel support from NATO.

---

This preprint was prepared with the AAS LaTeX macros v3.0.

Fig. 1.— Histograms of the distribution of the compactness parameters for identified galaxies and unidentified objects in three different magnitude bins. The three panels on the right shows the corresponding identification rates, the error bars are based on an assumed poissonian distribution.

Fig. 2.— Histograms of the parameter f which describes the effect of orientation on the amount of light entering a slit (f>1 means that the source main axis is preferentially oriented along the slit, f<1 indicates that it lies perpendicular to the slit). There is no evidence that the two populations (identified galaxies and unidentified objects) have different distributions.

Fig. 3.— The ratio R = N(galaxies in the top strips)/N(galaxies in the bottom strips), versus redshift. Redshift bins of 0.05 were adopted. Horizontal line shows the expected ratio, 174/418=0.416.

Fig. 4.— Finding chart for objects in the 0300+000 field. Eccentricities and orientations are those described in the text (section 3). Filled symbols represent objects for which spectra were obtained (see Table 1). North is at the top, East to the left.

Fig. 5.— A grey-scale image of the 0300+00 field.

Fig. 6.— Redshift histogram of the identified galaxies (confidence class $\geq 2$) in the 0300+00 field. Boxes represent relative numbers of stars and unidentfied objects (confidence class $\leq 1$.

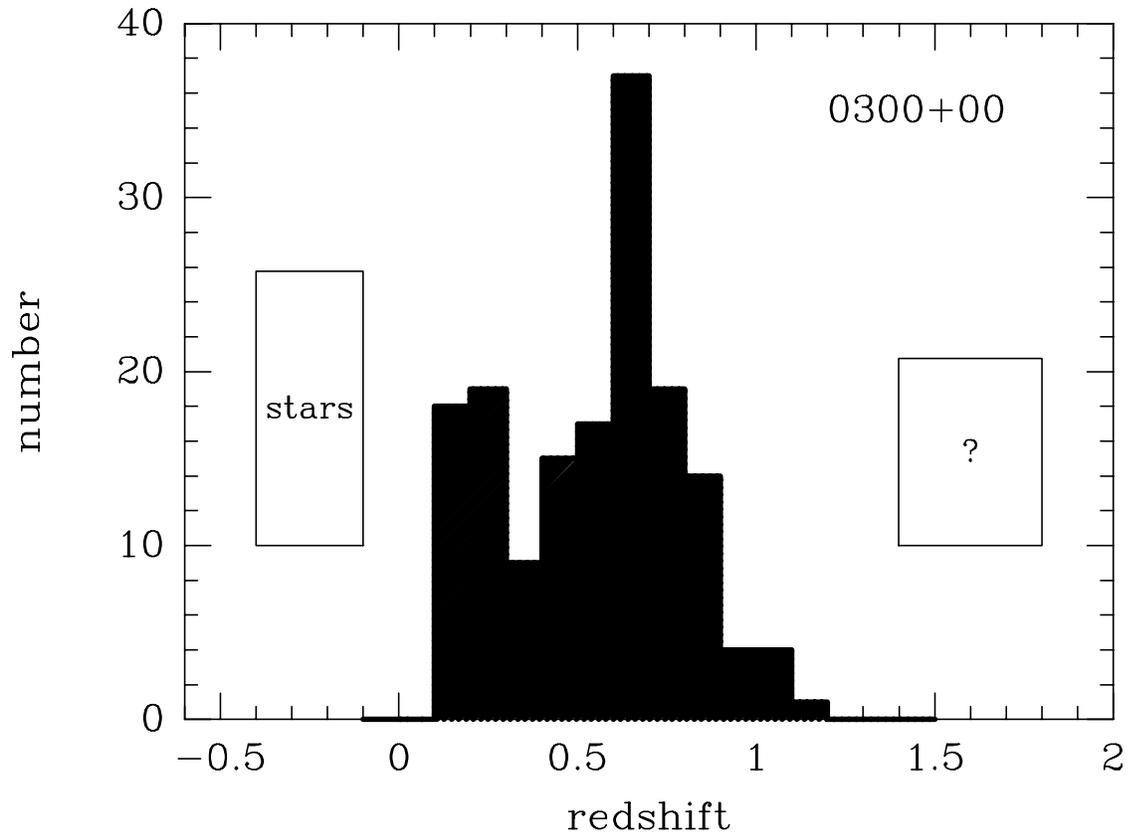